\documentclass[aps,twocolumn,showpacs,groupedaddress]{revtex4}%
\usepackage{amsmath}
\usepackage{graphicx}
\usepackage{amsfonts}
\usepackage{amssymb}%
\begin{document}
\title{Variational Approach to the Spin-boson Model With a Sub-Ohmic Bath}
\author{Hang Wong}
\author{Zhi-De Chen$^*$}
\affiliation{Department of Physics, Jinan University, Guangzhou
510632, China}

\begin{abstract}
The influence of dissipation on quantum tunneling in the
spin-boson model with a sub-Ohmic bath is studied by a variational
calculation. By examining the evolution of solutions of the
variational equation with the coupling strength near the phase
boundary, we are able to present a scenario of discontinuous
transition in sub-Ohmic dissipation case in accord with
Ginzburg-Landau theory. Based on the constructed picture, it is
shown that the critical point found in the general way is not
thermodynamically the critical point, but the point where the
second energy minimum begins to develop. The true cross-over point
is calculated and the obtained phase diagram is in agreement with
the result of numerical renormalization group calculation.

\end{abstract}
 \maketitle

\section{Introduction}
The spin-boson model is an important toy model for investigating
the influence of dissipation on quantum tunneling and has a wide
range of applications.\cite{leg,weiss}  Over decades the model has
been studied by various methods,\cite{leg} like path
integral,\cite{fey} renormalization group
calculation,\cite{bray,cha} variational
calculation,\cite{em,hew,zw,sh,chen,nie,ct} and the numerical
renormalization group(NRG) calculation,\cite{bulla} etc.. One
important issue is to study the cross-over from the delocalized to
localized phases as the dissipation increases. Most of the studies
are concentrated on the Ohmic dissipation case which is considered
as corresponding to real physical systems and the cross-over
picture is well understood. On the other hand, the situation for
the sub-Ohmic dissipation, which is of less physics interest but
still important for a well understanding of the spin-boson model,
has some confusions. Renormalization group calculation shown that
quantum tunneling is totally suppressed by dissipation for any
non-zero sub-Ohmic coupling at T=0,\cite{cha,leg} while different
conclusion was found by mapped the spin-boson model to an Ising
model\cite{sp} and using the well-known result for Ising
model.\cite{dys} The sub-Ohmic case was also studied by using
infinitesimal unitary transformation and the cross-over was found
to be discontinuous.\cite{keh} Recently, the NRG calculation,
which is considered as a powerful tool for investigation of the
Kondo model and its generalizations, confirmed the delocalized to
localized phase cross-over in sub-Ohmic dissipation case and the
cross-over is identified as continuous.\cite{bulla} Variational
calculation has been used to study the spin-boson model with a
Ohmic bath and the result of cross-over boundary is in good
agreement with the renormalization group
calculation.\cite{zw,sh,chen,nie} The variational calculation for
non-zero temperature\cite{sh} was generalized to sub-Ohmic case
recently and the discontinuous cross-over behavior was found to
exist at non-zero-temperature.\cite{ct} Up to now, the description
for this discontinuous cross-over is just limited to the
discontinuous change of the tunneling splitting at the cross-over
point, while a scenario for such a discontinuous cross-over is
still lacking. According to Ginzburg-Landau theory{\cite{gol,cl},
the evolution of the free energy around the critical point for the
first order(discontinuous) phase transition is rather complicated
and merely a discontinuous change of order parameter at the
cross-over point is certainly no enough for a complete description
of this discontinuous transition. In this paper, we present
further analysis on this discontinuous cross-over by examining the
evolution of the solutions of the self-consistent equation derived
from the variational calculation. It is found that the evolution
of the solutions near the phase boundary is consistent with the
general picture of the first order phase transition. Basing on the
constructed picture, it is shown that the critical points
determined in the general way are not thermodynamically critical
points and the true critical point is calculated. The arrangement
of the paper is as follows. In the next section, the model and a
brief explanation on variational calculation are presented. In
section III we present analysis on the discontinuous phase
transition by comparing the evolution of the solutions of the
self-consistent equation for Ohmic and sub-Ohmic dissipation cases
near the critical point. Conclusions and discussion are given in
the last section.

\section{The model and variational calculation}
 The Hamiltonian of the spin-boson model is given
by(setting $\hbar=1$)\cite{leg,weiss}
\begin{equation}
H=\frac{\epsilon}{2}\sigma_z+\frac{\Delta}{2}\sigma_x+\sum_k
b_k^{\dagger}b_k\omega_k+\sigma_z\sum_k c_k(b_k^{\dagger}+b_k),
\end{equation}
where $\sigma_i(i=x,y,z)$ is the Pauli matrix,
$b_k(b_k^{\dagger})$ is the annihilation(creation) operator of the
$k$th phonon mode with energy $\omega_k$ and $c_k$ is the coupling
parameter. The main interest of the present paper will be the zero
temperature so we set the bias $\epsilon=0$ in the following. It
is known that the solution of this model is  determined by the
so-called the bath spectral function(density) defined
as\cite{leg,weiss}
\begin{equation}
J(\omega)=\pi \sum_kc_k^2\delta(\omega-\omega_k).
\end{equation}
Generally $J(\omega)$ is characterized by a cut-off frequency
$\omega_c$ and has a power-law form, i.e.,
\begin{equation}
J(\omega)=\frac{\pi}{2}\alpha
\omega^s/\omega_c^{s-1},~~~~~~~0<\omega\le\omega_c,
\end{equation}
where $\alpha$ is a dimensionless coupling strength which
characterizes the dissipation strength. Parameter $s$ specifies
the property of the bath, $s=1$ is the  case of Ohmic dissipation
and $0\le s<1$ the sub-Ohmic dissipation case. It should be noted
that $J(\omega)$ can take some different forms,\cite{leg,ct} like
$J_1(\omega)=\frac{\pi}{2}\alpha
\omega^s/\omega_c^{s-1}e^{-\omega/\omega_c}$ and
$J_2(\omega)=\frac{\pi}{2}\alpha \omega^s/\omega_s^{s-1}$ with
$\omega_c \rightarrow \infty$,  while we find that the solution is
almost the same in sub-Ohmic case(see below).

As one can see from the Hamiltonian given in Eq.(1), when
$\Delta=0$  the localized phase is stable  since in this case we
have $[\sigma_z,H]=0$. This result implies that the coupling to
phonon bath alone cannot lead to tunneling and thus this problem
can be treated approximately in the way without coupling to the
bath as given in the quantum mechanics textbook.\cite{lan} To
ensure the tunneling is small which is a precondition of our
treatment, the following calculation is restricted to the
condition $\Delta/\omega_c\ll 1$. We denote the eigen-states of
spin-up(down)-plus-bath as
$|\uparrow\rangle|\phi_+\rangle(|\downarrow\rangle|\phi_-\rangle)$,
where $|\phi_{\pm}\rangle$ represent the eigen-state of the phonon
bath without tunneling. When the tunneling is taken into account,
the eigen-state of the whole system can be approximately given
by\cite{lan}
\begin{equation}
|\Phi_{\pm}\rangle=(|\uparrow\rangle|\phi_+\rangle\pm
|\downarrow\rangle|\phi_-\rangle)/\sqrt{2},
\end{equation}
then the tunneling splitting in the presence of dissipation is
\begin{equation}
\Delta'=\langle\Phi_+|H|\Phi_+\rangle-\langle\Phi_-|H|\Phi_-\rangle=
\Delta\langle\phi_+|\phi_-\rangle,
\end{equation}
a well known result that $\Delta'$ is determined by the overlap
integral of the phonon ground states.\cite{chen,nie} In the
absence of tunneling(i.e., $\Delta=0$), Hamiltonian (1) can be
diagonalized by a well-known displaced-oscillator-transformation
and the phonon ground states are the so-called
displaced-oscillator-states\cite{mah}
$$
|\phi_{d\pm}\rangle={\rm
exp}\{\pm\sum_k\frac{c_k}{\omega_k}(b_k-b_k^{\dagger})\}|0\rangle,
$$
where $|0\rangle$ is the vacuum state of the phonon. The essence
of the variational calculation is that, in the presence of
tunneling, the phonon ground states are suggested to still have
the same from, i.e.,
\begin{equation}
|\phi_{\pm}\rangle={\rm
exp}\{\pm\sum_kg_k(b_k-b_k^{\dagger})\}|0\rangle,
\end{equation}
but leaving the parameter $g_k$ to be determined from the
condition that the ground state energy of the whole system is a
minimum with respect to $g_k$. Substituting the above equation to
Eq.(4), the ground state energy of the whole system is found to be
\begin{equation}
E[g_k]=\sum_k(\omega_kg_k^2-2c_kg_k)-\frac{1}{2}\Delta\exp\{-2\sum_kg_k^2\},
\end{equation}
which is a functional of $g_k$, then $\frac{\delta E}{\delta
g_k}=0$ leads to
\begin{equation}
g_k=\frac{c_k}{\omega_k+ \Delta\exp\{-2\sum_kg_k^2\}},
\end{equation}
the tunneling splitting, by Eq.(5), is given by
\begin{equation}
\Delta'= K \Delta, ~~~~~~K=F[g_k]\equiv \exp\{-2\sum_kg_k^2\}.
\end{equation}
Using Eq.(8) and the definition of the spectral function, we find
that $K$ is determined by the following self-consistent(or
variational) equation
\begin{equation}
K=f(K),~~~~~~f(K)\equiv {\rm exp}
\{-\alpha\int_0^1\frac{x^s~dx}{[x+(\Delta/\omega_c)K]^2} \}.
\end{equation}
Such a kind of self-consistent equation has been derived in
previous works\cite{hew,zw,sh,chen,nie,ct} and it plays an
important role in dealing with the cross-over from the delocalized
to localized phases. It is easy to see that $K=0$ is the trivial
solution of Eq.(10) and this solution represents the localized
phase. When the coupling strength $\alpha$ is large enough, $K=0$
is the only solution of the self-consistent equation, while as
$\alpha$ decreases to some value $\alpha_c$, the self-consistent
equation begins to have, in addition to the trivial solution,
non-zero solutions, then $\alpha_c$ is identified as the critical
point at where the cross-over from the localized($K=0)$ to
delocalized($K>0$) phases happens. This is the general way to
determine the phase boundary used in previous
works.\cite{zw,sh,chen,nie,ct}
\begin{figure}
\centering
\includegraphics[width=0.48\textwidth]{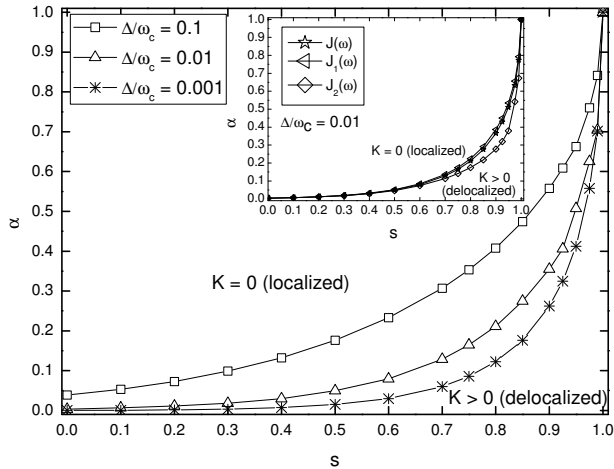}
\caption{Phase boundary determined by $\alpha_c$ for various
$\Delta/\omega_c$. The inset shows the comparison with the result
by using different spectral function $J_1(\omega)$ and
$J_2(\omega)$(see the text) in the case of
$\Delta/\omega_c=0.01$.}
\end{figure}

In the case of Ohmic dissipation, i.e., $s=1$, the self-consistent
equation can be solved analytically and the phase boundary is
found to be $\alpha_c=1$ in the case of $\Delta/\omega_c\ll 1$, in
agreement with the renormalization group
calculation.\cite{zw,nie,sh} In the case of sub-Ohmic dissipation,
the self-consistent equation can be solved numerically and the
phase boundary between the localized($K=0)$ and delocalized($K>0$)
phases determined in this way is shown in Fig.1. The result by
using different spectral functions, i.e., $J_1(\omega)$ and
$J_2(\omega)$, are also shown in the inset. Our result shows that,
for $\Delta/\omega_c\ll 1$, the phase boundary is almost the same
for all three spectral functions as $s\le 0.7$, while $\alpha_c$
is a little bit lower for $J_2(\omega)$ when $s>0.7$. Also, it is
found that the relation between the critical coupling $\alpha_c$
and $\Delta/\omega_c$ has a simple power-law form $\alpha_c\propto
(\Delta/\omega_c)^{1-s}$ as found by NRG calculation.\cite{bulla}
Notably, such a relation can be deduced from Eq.(16) in
ref.\cite{keh} by using the spectral function given
here.\cite{note} However, as we shall show in the next section,
the $\alpha_c$ determined in this way for sub-Ohmic case is not
thermodynamically the critical point, but just the the limit of
metastability for superheating of the first order phase
transition.\cite{cl}


\section{The discontinuous cross-over in sub-Ohmic
case} Now we turn to present a scenario for such a discontinuous
cross-over  from the delocalized to localized phases in sub-Ohmic
case. The key point is to examine the evolution of the solutions
of the self-consistent equation with the coupling strength
$\alpha$ near the phase boundary. For clarity, we first see what
happens in the Ohmic case. Fig.2 shows the evolution of the
solutions of Eq.(10) with the increase of $\alpha$ in Ohmic
dissipation case. When $\alpha>\alpha_c$, we have the trivial
solution only, while a non-zero solution($K_1\not=0$) appears for
$\alpha<\alpha_c$. As one can see from the figure, the non-zero
solution $K_1$ continuously tends to 0 as $\alpha$ approaches
$\alpha_c$. This is consistent with the picture of a
continuous(second order) transition: \cite{gol,cl} above the
critical point($\alpha>\alpha_c$), there is only one stable
phase(one energy minimum located at some $g_{k0}$ satisfying
$F[g_{k0}]=0$ in the present case), below the critical point, this
stable phase becomes unstable($E[g_{k0}]$ becomes the maximum of
the energy) and a second stable phase appears($E[g_{k1}]$ is the
new energy minimum, where $F[g_{k1}]=K_1>0$), the cross-over
behavior is continuous.

The situation for the sub-Ohmic dissipation case is qualitatively
different. As shown in Fig.3, when $\alpha<\alpha_c$,
\begin{figure}
\centering
\includegraphics[width=0.48\textwidth]{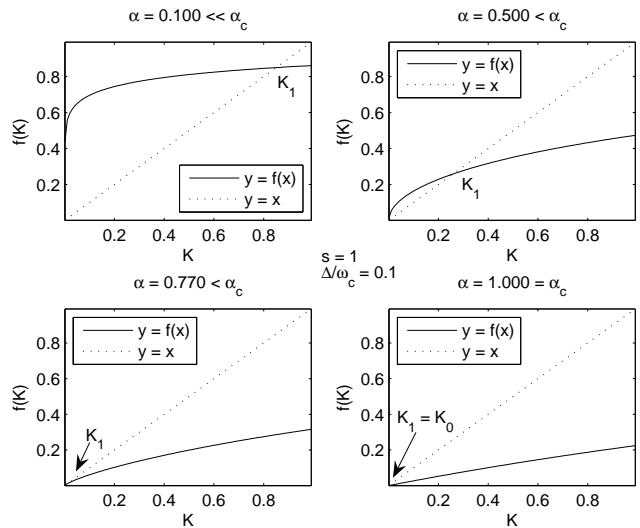}
\caption{Evolution of the solutions of Eq.(10) with the increase
of coupling strength $\alpha$ for $\Delta/\omega_c=0.1$ in the
case of Ohmic dissipation $s=1$.}
\end{figure}
\begin{figure}
\centering
\includegraphics[width=0.48\textwidth]{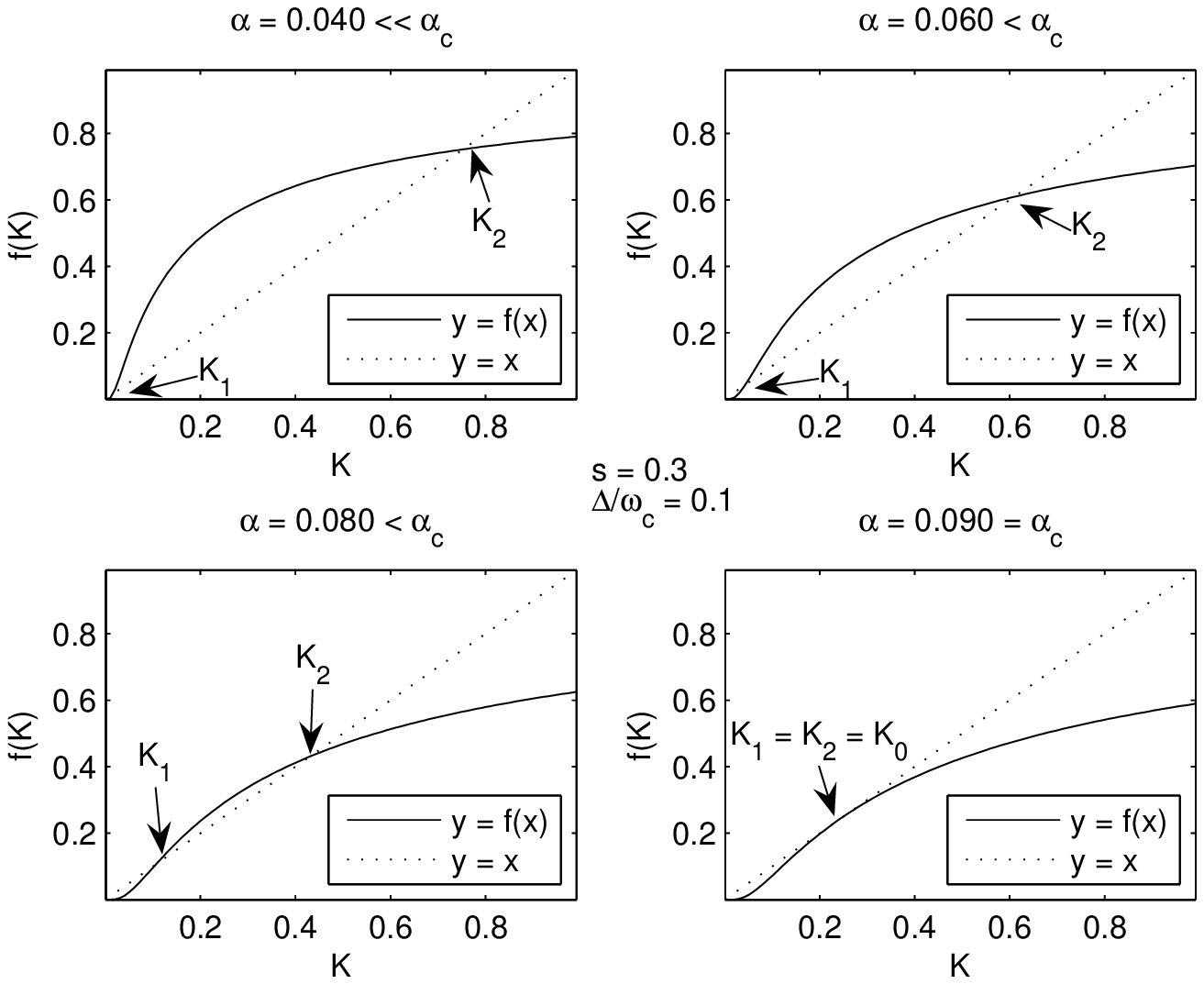}
\caption{Evolution of the solutions of Eq.(10) with the increase
of coupling strength $\alpha$ for $\Delta/\omega_c=0.1$ in the
case of sub-Ohmic dissipation $s=0.3$. }
\end{figure}
\begin{figure}
\centering
\includegraphics[width=0.23\textwidth]{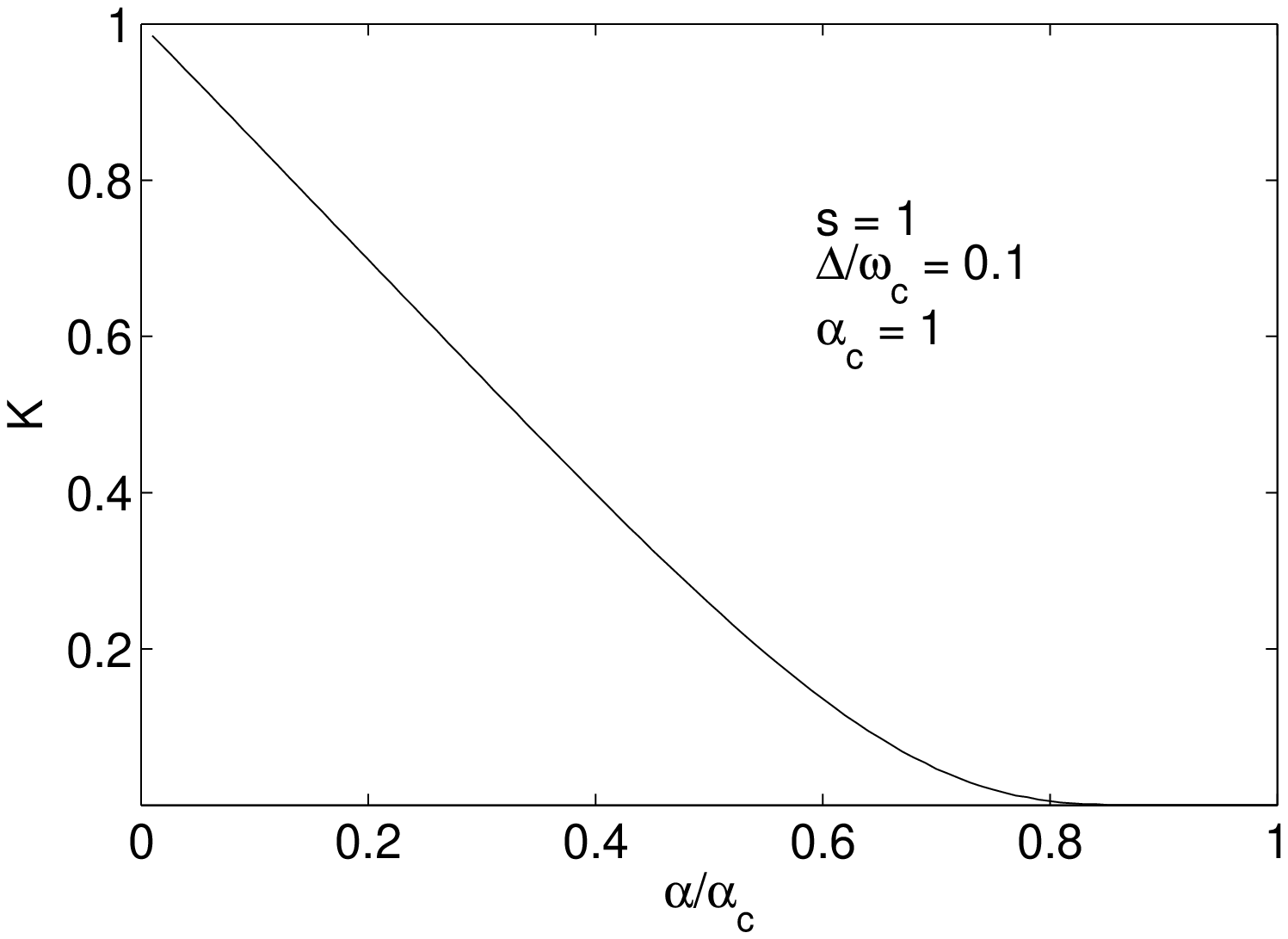}
\includegraphics[width=0.23\textwidth]{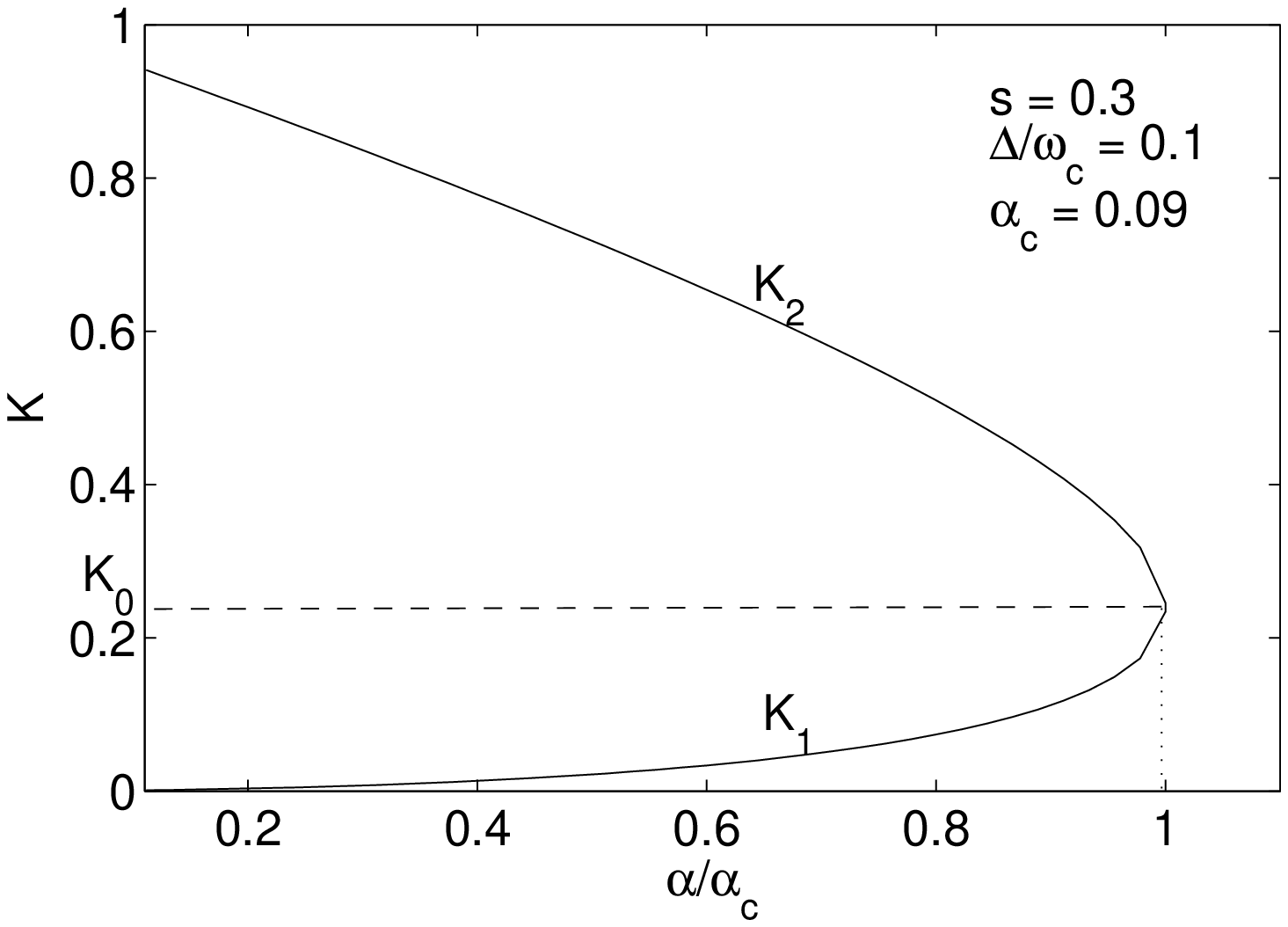}
\caption{The $\alpha$-dependence of the non-zero solution of
Eq.(10) for $s=1$(left) and $s=0.3$(right). As
$\alpha/\alpha_c\rightarrow 1$, the non-zero solution of $s=1$
approaches 0 continuously, while for $s=0.3$, the non-zero
solution jumps from $K_0\not=0$ to 0.} \label{aaa}
\end{figure}
\begin{figure}
\centering
\includegraphics[width=0.48\textwidth]{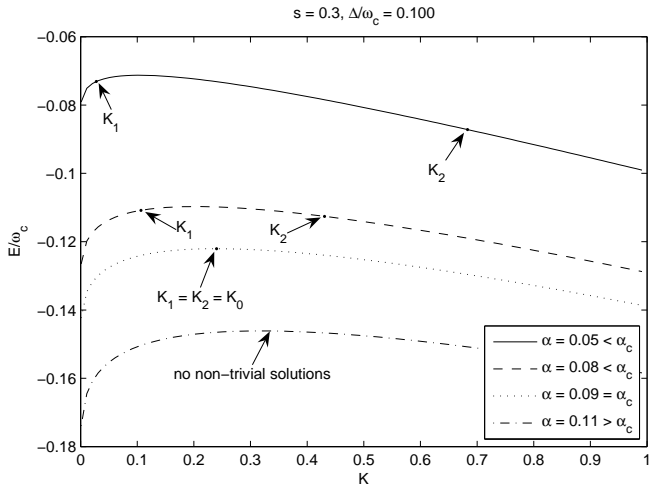}
\caption{Evolution of the K-dependence of the ground state energy E
with coupling strength $\alpha$ for sub-Ohmic case($s=0.3$) when
$\Delta/\omega_c=0.1$. }
\end{figure}
\begin{figure}
\centering
\includegraphics[width=0.48\textwidth]{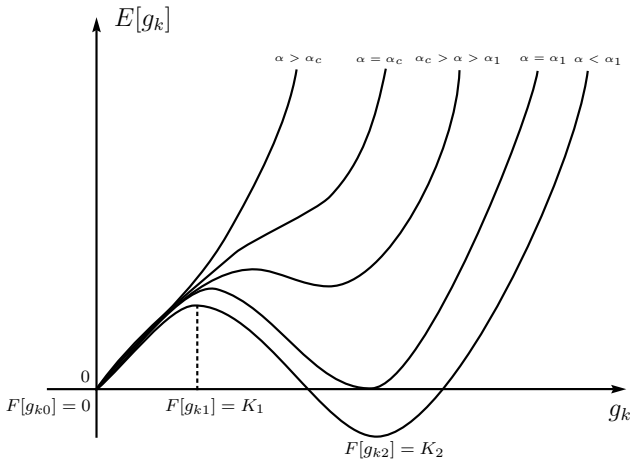}
\caption{The evolution of the energy extrema with the coupling
strength $\alpha$ for the sub-Ohmic case. Thermodynamically the
cross-over point is $\alpha_1$, while $\alpha_c$ is just the point
where the second minimum begins to develop.}
\end{figure}
\begin{figure}
\centering
\includegraphics[width=0.48\textwidth]{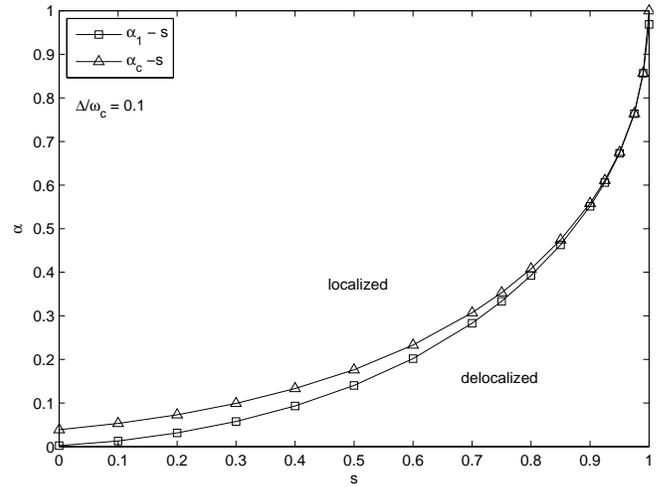}
\caption{The phase boundary determined by $\alpha_c$ and
$\alpha_1$ in the case of $\Delta/\omega_c=0.1$. }
\end{figure}
there are {\it two} non-zero solutions of Eq.(10)($K_2>K_1\not=0$)
in additional to the trivial solution.\cite{e1}  As the coupling
strength $\alpha$ increases, $K_2$ decreases while $K_1$ increases
and tends to meet $K_2$ as $\alpha$ approaches $\alpha_c$. At
$\alpha=\alpha_c-0$, we have $K_1=K_2=K_0\not=0$ and at this point
$K_0$ is the point of tangency for the line $y=x$ and curve
$y=f(x)$. At $\alpha=\alpha_c+0$, the solution $K_0$ disappears
{\it suddenly} and only the trivial solution is found. The
$\alpha$-dependence of the non-zero solutions of Eq.(10) for Ohmic
and sub-Ohmic dissipation cases are shown in Fig.4 where one can
see the qualitatively different cross-over behavior. The result of
sub-Ohmic dissipation case clearly shows that the cross-over is
discontinuous since the non-zero solution of Eq.(10) and thus the
tunneling splitting $\Delta'$ changes discontinuously at the point
$\alpha=\alpha_c$. Such a behavior was found before\cite{keh,ct}
and took as the evidence for a discontinuous transition since the
tunneling splitting has a physics meaning of the order parameter.

What we want to emphasize here is the  two non-zero solutions when
$\alpha<\alpha_c$. Physically we need to know which solution is
stable and the meaning of the second non-zero solution. As one can
see from Fig.4(right), the non-zero solution $K_1$ increases with
$\alpha$, one can intuitively conclude that $K_1$ is unstable
since physically the tunneling splitting should decrease with
$\alpha$. Basing on the variational calculation, we cannot make
further analysis on the stability of the solution, so we turn to
energy analysis, i.e., to see which solution is energy preferable.
Typical evolution of the $E-K$ curve with $\alpha$ is shown in
Fig.5. The result shows that, we have $E(K_1)>E(0)$ and
$E(K_1)>E(K_2)$ when $\alpha\ll \alpha_c$, $E(0)$ decreases while
both $E(K_1)$ and $E(K_2)$ increase relatively as $\alpha$
increases but $E(K_1)$ is always the largest, finally
$E(K_1)=E(K_2)=E(K_0)$ as $\alpha=\alpha_c$ and we have
$E(0)<E(K_0)$. This implies that, both $K=0$ and $K_2$ are energy
preferable while $K_1$ is unstable when $\alpha<\alpha_c$. Such a
result is consistent with the scenario of a first order
transition. In the scenario of the first order
transition,\cite{gol,cl} below the critical point, there are two
free energy minima and a maximum lies between, while above the
critical point, only one global free energy minimum survives. In
the sub-Ohmic dissipation case, when $\alpha<\alpha_c$, three
solutions of Eq.(10) represent the two energy minima and one
energy maximum, that is, $E[g_{k0}]$ with $F[g_{k0}]=0$ and
$E[g_{k2}]$ with $F[g_{k2}]=K_2$ are the two energy minima, while
$E[g_{k1}]$ with $F[g_{k1}]=K_1$ is the energy maximum lies
between as shown in Fig.6. As $\alpha$ increases and approaches
$\alpha_c$, $E[g_{k1}]$ tends to meet $E[g_{k2}]$ and at the point
$\alpha=\alpha_c$, these two energy extrema merge into a point of
inflection at $F[g_k]=K_0$, then only one energy minimum
$E[g_{k0}]$ survives when $\alpha>\alpha_c$. Based on the picture
for the discontinuous phase transition, it is now clear that
$\alpha_c$ is {\it not} the critical point for the cross-over to
happen, but just the point where the second energy minimum begins
to develop. $\alpha_c$ can be considered as the limit of
metastability for superheating,\cite{cl} i.e., the limit of
metastability for increasing the dissipation strength in the
present case. Thermodynamically the critical point, as shown in
Fig.6, should be $\alpha_1$ where we have\cite{gol,cl}
\begin{equation}
E(K_2)=E(0),
\end{equation}
from this, $\alpha_1$ can be determined by Eqs.(7), (8) and (10).
 Comparison between the phase boundary determined by
$\alpha_c$ and $\alpha_1$ is shown in Fig.7. It is easy to see
that $\alpha_1<\alpha_c$ while the difference between $\alpha_c$
and $\alpha_1$ decreases as $s$ increases and tends to zero as
$s\rightarrow 1$ where the transition becomes continuous. We also
find that the difference between $\alpha_c$ and $\alpha_1$
decreases with $\Delta/\omega_c$.
\begin{figure}
\centering
\includegraphics[width=0.48\textwidth]{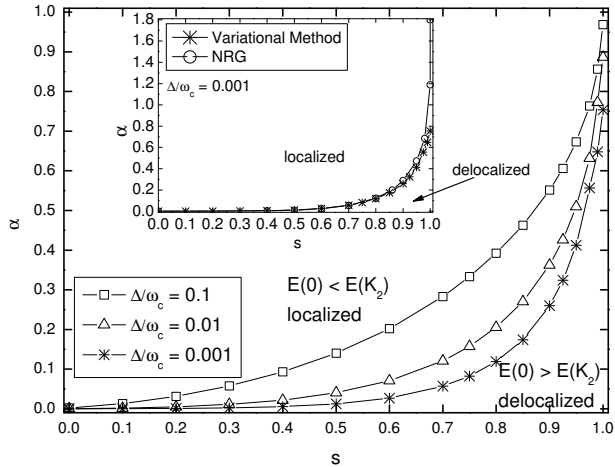}
\caption{The phase boundary determined by $\alpha_1$ for various
$\Delta/\omega_c$. The inset shows the comparison with the NRG
calculation in the case of $\Delta/\omega_c=0.001$.}
\end{figure}
The phase boundary deduced in this way is shown in Fig.8 which is
similar to that shown in Fig.1 but with all the critical points
lower. It is found that the phase boundary determined by
$\alpha_1$ is in good agreement with that obtained by NRG
calculation when $\Delta/\omega_c \le 0.01$.

\section{Conclusions and discussion}
In conclusion, we have study the cross-over behavior from
localized  to delocalized phases of a spin-boson model with a
sub-Ohmic bath by variational method.  By examining the evolution
of the solutions of self-consistent equation (10) with the
coupling strength, we are able to present the scenario of the
discontinuous transition in sub-Ohmic dissipation case. Based on
the constructed picture, it is shown that the $\alpha_c$, at where
the self-consistent equation begins to have non-zero solutions, is
not thermodynamically the critical point, but just the point where
the second energy minimum begins to develop. The true critical
point is determined according to Ginzburg-Landau theory for the
first order phase transition and the obtained phase boundary  is
in agreement with the NRG calculation. Our analysis shows that the
cross-over behavior in spin-boson model is directly related to the
evolution of solutions of the self-consistent equation derived
from the variational calculation. The evolution behavior of
solutions for a continuous cross-over(in Ohmic dissipation case)
is qualitatively different from that of a discontinuous
cross-over(in sub-Ohmic dissipation case). The present work, on
one hand, provides convincing evidence for a discontinuous
cross-over in sub-Ohmic case and on the other hand, demonstrates
the new way to deal with the cross-over behavior in spin-boson
model by the variational method.

According to the definition of stable and unstable fixed points
for renormalization group,\cite{jp} geometrically one can see from
Fig.3 that, both $K=0$ and $K_2$ are stable fixed points while
$K_1$ is unstable fixed point as $\alpha<\alpha_c$ in sub-Ohmic
case. On the other hand, we only have one stable fixed point(i.e.,
$K_1$) and one unstable fixed point as $\alpha<\alpha_c$ in Ohmic
case. This result is in agreement with the NRG calculation,
\cite{bulla} where 3 fixed points(2 stable and 1 unstable) were
found in sub-Ohmic case while the third unstable fixed point
disappeared in Ohmic case. However, the cross-over behavior in
sub-Ohmic case was identified as continuous in NRG calculation,
this implies further analysis is needed for seeking a deeper
relation. Although the work by Kehrein and  Mielke is not based on
the variational calculation,\cite{keh} the cross-over behavior was
studied by a self-consistent equation and the discontinuous
behavior was judged by the discontinuous change of the tunneling
splitting at the critical point $\alpha_c$,  where the
self-consistent equation begins to have non-zero solutions. Some
results, like the $(\Delta/\omega_c$) dependence of critical
coupling $\alpha_c$ and the $s$-dependence of tunneling splitting
at the critical point also show quantitative agreement with our
work determined from Eq.(10) at $\alpha=\alpha_c$. This may lead
to a conclusion the the critical point determined in
ref.\cite{keh} is just $\alpha_c$ given in the present work, i.e.,
not thermodynamically the critical point.

The author(Chen) thanks Dr. N. H. Tong for supplying the data of
phase diagram by NRG  calculation. This work was supported by a
grant from the Natural Science Foundation of China  under Grant
No. 10575045.

 $^*$ Author to whom correspondence should be addressed. E-mail: tzhidech@jnu.edu.cn


\begin{thebibliography}{99}
\bibitem {note} The expression deduced from Eq.(16) in
ref.\cite{keh} has an additional $s$-dependent factor, but
numerical analysis shows that this addition factor is of less
importance for $0<s<1$.
\bibitem {e1} The existence of two non-zero solutions for
variational equation when $\alpha<\alpha_c$ can also be deduced
from the curves shown in Fig.1 of ref.\cite{ct}.
\end{thebibliography}
\end{document}